\documentclass[aps, pre, twocolumn,superscriptaddress,amsmath,amssymb,floatfix]{revtex4-1} 

\newif\ifpdf\ifx\pdfoutput\undefined\pdffalse\else\pdfoutput=1\pdftrue\fi

\newcommand{\effe}{{\mathcal{E}}} 

\usepackage{graphicx}
\usepackage[usenames]{color} 
\usepackage{comment} 
\usepackage{bm}
\newcounter{abc}

\newcommand{\be}{\begin{equation}} 
\newcommand{\ee}{\end{equation}}
\newcommand{\bea}{\begin{eqnarray}} 
\newcommand{\eea}{\end{eqnarray}}

\begin{document}

\title{Accurate simulation estimates of phase behaviour in ternary mixtures with prescribed composition}
\author{Nigel B. Wilding} 


\affiliation{Department of Physics, University of Bath, Bath BA2
7AY, United Kingdom.}

\begin{abstract}

This paper describes an isobaric semi-grand canonical ensemble Monte
Carlo scheme for the accurate study of phase behaviour in ternary fluid
mixtures under the experimentally relevant conditions of prescribed
pressure, temperature and overall composition. It is shown how to tune
the relative chemical potentials of the individual components to target
some requisite overall composition and how, in regions of phase
coexistence, to extract accurate estimates for the compositions and
phase fractions of individual coexisting phases. The method is
illustrated by tracking a path through the composition space of a
model ternary Lennard-Jones mixture.

\end{abstract}

\maketitle

\section{Introduction} 

Fluids comprising multiple distinct components or `species' are
pervasive in chemistry, physics and chemical engineering where they
feature in contexts as diverse as chemical solutions, fuels,
lubricants, microemulsions and cleaning products. Understanding the
phase behaviour of such mixtures is important for exploiting their novel
properties, for example, by ensuring a lubricant remains fluid under
prescribed working conditions, or in designing processes for separation
of hydrocarbons. 

In principle, the phase behaviour of a fluid mixture can be determined
experimentally. However, the parameter space of composition, temperature
and pressure is typically large, rendering this costly in both time and
money. Accordingly, there is much interest in efficiently predicting the
phase behaviour of multicomponent fluids using computer simulation. In
attempting to do so, it is reasonable to align oneself with the
experimental situation in which (typically) one fixes all but one or two
of the system parameters and varies the remaining ones within some
bounds. Thus, for instance, in a ternary mixture one might set the
temperature and pressure, and vary the composition of the mixture in a
prescribed way, eg. by varying the relative concentration of a pair of
components. Thus one explores the phase behaviour along a particular
{\em path} in composition space. 

In this paper we shall consider a computational strategy for determining
what happens if such a pathway intersects a region of phase coexistence.
When this occurs the system will separate into two or more phases, and
we seek to determine --accurately-- their compositions and relative
quantities accurately for all points on the path.

\section{Phase separation and fractionation}

The overall composition of a three component (ternary) mixture is specified
in terms of the set of concentrations of the individual components,
$x_1^{(0)}, x_2^{(0)},x_3^{(0)}$ or, more succinctly, $\{x_i^{(0)}\}$ with
$i=1\ldots 3$. Here 

\be
x_i^{(0)}=N_i/N\;,
\ee
with $N_i$ the total number of particles of species $i$ in the system and $\sum_{i=1}^3
N_i$ the total number of particles. Clearly
overall concentrations sum to unity, ie. $\sum_{i=1}^3x_i^{(0)}=1$, so
that only two concentrations need be specified to define the composition.

If for some prescribed overall composition, temperature $T$, and
pressure $p$, the system phase separates, the composition of the
coexisting phases will, in general, differ from the overall
composition--a phenomenon termed `fractionation'. However conservation
of particles implies that the concentrations of the individual phases
are related to the overall composition by a generalized lever rule. Let
us assume that there are $m$ coexisting phases, then

\be
x_i^{(0)}=\sum_{\gamma=1}^m \xi^{(\gamma)}x_i^{(\gamma)}\:.
\ee
Here $x_i^{(\gamma)}$ is the concentration of component $i$ in phase $\gamma$, defined as

\be
x_i^{(\gamma)}=N_i^{(\gamma)}/\sum_{i=1}^3 N_i^{(\gamma)}\;,
\ee
while $\xi^{(\gamma)}$ is the ``phase fraction'' of phase $\gamma$ ie. the fraction of all particles
that are found in phase $\gamma$:

\be
\xi^{(\gamma)}=\frac{\sum_{i=1}^3 N_i^{(\gamma)}}{N}\,.
\ee
Of course phase fractions sum to unity, so $\sum_{\gamma=1}^m \xi^{(\gamma)}=1$.  

It follows that in order to specify the coexistence properties of the
system at some $\{x_i^{(0)}\}$, one must determine $\{x_i^{(\gamma)}\}$
and $\xi^{(\gamma)}$ for each phase present. Before describing how this
determination can be made, we shall expand on the concept of a path in
the space of overall composition.

\section{Phase diagram pathways}

\begin{figure}[h]
\includegraphics[type=pdf,ext=.pdf,read=.pdf,width=0.85\columnwidth,clip=true]{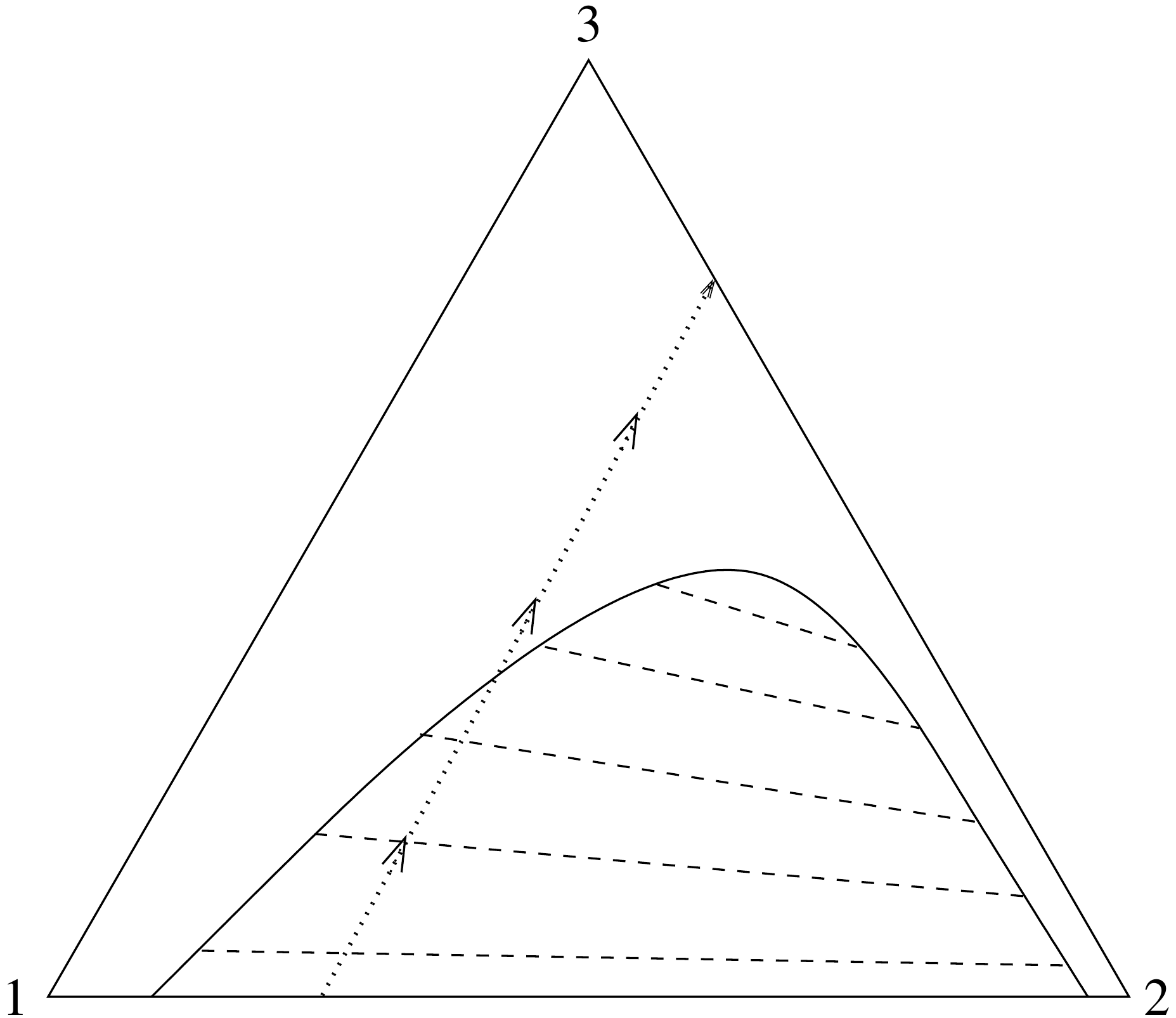}\\

\caption{Schematic of an exemplary phase diagram of a ternary mixture at
fixed temperature and pressure, represented in terms of a Gibbs phase
triangle as described in the text.}
\label{fig:schem_trian} 
\end{figure}

Let us illustrate what we mean by a phase diagram pathway in the context of
a ternary mixture. At fixed temperature and pressure a possible phase
diagram for such a system is shown in Fig.~\ref{fig:schem_trian} in the
form of a Gibbs phase triangle \cite{West1982}. 
The diagram includes a region of two phase coexistence with tie lines
that shrink to zero at a critical point. These tie lines connect points
that represent the compositions of the coexisting phases. Any
point of overall composition within the coexistence region will divide
a tie line into two segments, the relative lengths of which 
yield the phase fractions of the coexisting phases.

Also shown in Fig.~\ref{fig:schem_trian} is an experimental pathway in
composition space that passes through the coexistence region (dotted
line). Existing simulation approaches are not well suited to
determining coexistence properties along such a path. Instead they are
tailored to the task of determining the boundary of the coexistence
region itself and the associated tie lines (see eg.
\cite{Tsang1995,Escobedo1999,Errington2005}). Of course, from this
information one can in principle {\em infer} the coexistence properties
(ie. the compositions and phase fractions) along the experimental path by
graphical means--though the procedure may be of questionable accuracy.
However, such a graphical solution is impracticable for mixtures of more
than three components because compact representations of phase behaviour
for such systems do not exist. Moreover, even for binary and ternary
systems, a straightforward graphical procedure is only feasible when
both pressure and temperature are fixed: if the experimental path
involves changes in temperature or pressure in addition to compositional
changes, it becomes necessary (with existing simulation methods) to map
a portion of the full phase coexistence {\em surface} in order to infer
the behaviour along the (one-dimensional) path of interest, ie. one
needs to solve a higher dimensional problem than the one of actual
interest. 


Thus it is of interest to develop new simulation strategies that are
potentially more direct, accurate and flexible that those currently
available and which may generalize to multicomponent mixtures and
arbitrary paths through the phase diagram. In the current paper we take a
step towards this goal by showing how one can track an arbitrary path in
the space of overall composition of a ternary mixture.

\section{Existing simulation approaches}

Let us first briefly summarize principal existing simulation
methodologies for obtaining the phase behaviour of ternary fluids. Two
phase coexistence in a Lennard-Jones mixture was studied by Escobedo
using Gibbs-Duhem integration (GDI) in the semi-grand canonical
ensemble~\cite{Escobedo1999}. GDI requires independent knowledge of a
coexistence state point to bootstrap the integration, but once going it
follows the coexistence binodal through composition space. Whilst this
delivers more or less complete information on the phase behaviour
(modulo integration errors), it doesn't give direct information on the
coexistence properties (ie. the $\{x_i^{(\gamma)}\}$ and
$\xi^{(\gamma)}$) along a particular experimentally relevant path in the
phase diagram. Instead this behaviour has to be inferred from the
knowledge of the binodal and the tie lines, as described in the previous
section.

Errington \cite{Errington2005} has described both a semi-grand canonical
and a full grand canonical Monte Carlo (MC) scheme for studying fluid
phase behaviour in multicomponent mixtures. This is a potentially
powerful approach, aspects of which we shall draw upon below.
Nevertheless, it too is tailored for determining the coexistence binodal
itself and without the extensions that we shall describe, is ill-suited
to the task of tracking a particular experimental path and accurately
determining the associated coexistence properties.

In this respect Gibbs ensemble Monte Carlo (GEMC) is more flexible,
since one can simply adjust the overall densities of the components.
However, as soon as the path to be followed passes close to the
coexistence boundary (so that one of the phase fractions becomes small),
the volume of one of the GEMC boxes also becomes small and pronounced
finite-size effects are to be expected. The method can therefore only
provide reliable results for state points that are well within the
coexistence region, leading to the same problems experienced by GDI and
Errington's approach. The constant pressure GEMC technique has been
applied by Tsang {\em et al} \cite{Tsang1995} to obtain the coexistence
properties of the same ternary mixture that was studied by Escobedo
\cite{Escobedo1999}.  

Finally we mention a more bespoke technique for studying phase behaviour
in mixtures, namely the pseudo ensemble method of Vrabec and
Hasse\cite{VRABEC2002}. This finds coexistence properties for prescribed
choices of temperature and the composition of the liquid, which allows
for calculations of the binodal, but does not seem capable of
determining phase coexistence properties for prescribed values of the
{\em overall} composition.

\section{Method}

\subsection{Isobaric semi-grand canonical ensemble}
\label{sec:SGCE}

We turn now to the present method, which is in fact an adaptation of a fully
grand canonical MC approach originally developed for estimating cloud
and shadow points in the phase diagrams of polydisperse fluids
\cite{Buzzacchi2006,Wilding2008}. Here, however, we shall work within the
isobaric semi-grand canonical ensemble (SGCE). Within this framework the
system volume $V$, the energy, and the instantaneous number densities of
the species all fluctuate \cite{Kofke1988}, while the particle number
$N$, pressure $p$, temperature $T$, and a set of chemical potentials
$\{\mu_i\}$ are all prescribed. The latter are measured with respect to
an arbitrarily chosen reference species, ie. for a ternary mixture there
are actually only two independent chemical potential {\em differences}. 

Operationally, the sole difference between the isobaric,
semi-grand-canonical ensemble and the familiar constant-$(N,p,T)$ ensemble
\cite{Frenkelsmit2002} is that one implements MC updates that select a
particle at random and attempts to change its species label $i$, say, to another species
$i^\prime$, chosen randomly. This proposal is accepted or rejected with a Metropolis
probability controlled by the change in the internal energy and the
chemical potential \cite{Kofke1987}. This probability reads:

\[
p_{\rm acc}={\rm min}\left[1,\exp{(-[\Delta \Phi+\mu_i-\mu_{i^\prime}]/T)}\right]\:,
\]
where $\Delta \Phi$ is the internal energy change associated with the
relabeling operation. 

\subsection{Tracking a composition space pathway}

The SGCE is the appropriate ensemble for studying phase coexistence in
multicomponent mixtures for two reasons. Firstly it realizes the common
experimental scenario of fixed pressure and temperature; and secondly it provides for
fluctuations in the densities of the individual species on the scale of
the system size. This latter feature is crucial to allow for separation
into differently fractionated phases. However, in order to track
a composition space pathway, one first needs to bootstrap the tracking
procedure by determining the set of chemical potentials that target {\em
some} state point on the pathway. In what follows we describe how this
can be achieved straightforwardly in two situations, namely (i)
when the pathway passes through a one phase region of the phase diagram, or (ii)
it intersects an axis of the ternary phase diagram, ie. coincides with the
limit of a binary system.

\subsubsection{Bootstrapping the method in the one phase region}

The task is to determine, for some given $T$ and $p$, the set of
chemical potentials $\{\mu_i\}$ that yield some prescribed set of
``target'' concentrations $\{x_i^{(0)}\}$. By this we mean that the ensemble
averages of the fluctuating overall concentrations $\{\bar x_i\}$ matches
$\{x_i^{(0)}\}$. In a single phase region, the non-equilibrium potential
refinement (NEPR) scheme \cite{Wilding2003a} enables the efficient
iterative determination of $\{\mu_i\}$ from a single simulation run, and
without the need for initial guesses. To achieve this, the $\{\mu_i\}$
are continually updated (in the course of a simulation run) in such a
way as to minimize the deviation of the instantaneous concentrations
$\{x_i\}$ from the target.  This procedure realizes a non-equilibrium
steady state for which $\{\bar x_i\}=\{x_i^{(0)}\}$. However, since tuning
model parameters  `on-the-fly' in this manner violates detailed balance,
the $\{\mu_i\}$ thus obtained is not the equilibrium solution one seeks.
Nevertheless by performing a series of iterations in which the degree of
modification made to $\{\mu_i\}$ at each step is successively reduced,
one can drive the system towards equilibrium, thereby obtaining the
correct $\{\mu_i\}$.  Full details of the procedure are provided in
\cite{Wilding2003a}. An alternative scheme for determining chemical
potentials has recently been described by Malasics {\em et
al}~\cite{Malasics2008}.

Once $\{\mu_i\}$ has been determined for a state point on the
composition space path of interest, one can step along this path using
histogram extrapolation techniques \cite{ferrenberg1989}, without having
to apply the NEPR technique again. The basic idea is measure fluctuating
quantities at the initial state point on the path and reweight them to
determine the $\{\mu_i\}$ corresponding to a nearby state point on the
path. The resulting $\{\mu_i\}$ are used to perform a fresh SGCE
simulation at the new state point, and the process is repeated.

\subsubsection{Tracking the path into a two phase region}
\label{sec:tracking}

Suppose now that the path being followed through composition space
enters a two phase region. The boundaries of this region can be located
via extended sampling. The idea is to employ standard Multicanonical
biasing techniques \cite{berg1992} to extend the range sampled by some
order parameter. The precise choice of order parameter is not crucial,
what matters is that it distinguishes clearly between the coexisting
phases. In the case of a liquid-vapour transition, the system volume $V$
is a suitable order parameter and we monitor the order parameter
probability distribution function (pdf), $P(V)$. Close to the coexistence
boundary this pdf exhibits a second much smaller peak representing
contributions from the incipient phase.  The appearance of the new phase
means that fractionation will start to occur and in order to remain on
the desired path in overall composition space, one needs to account for
this. To do so involves determining values for the $\{\mu_i\}$ that
yield the prescribed overall composition across the coexisting phases.
At the same time, one wishes to determine  accurately the
$\{x_i^{(\gamma)}\}$ and $\xi^{(\gamma)}$, notwithstanding the fact that
the relative quantity of the incipient phase is close to zero.

All this can be achieved via the following strategy. For given choices
of $\{x_i^{(0)}\}$ , $T$, and $p$, one tunes
$\{\mu_i\}$ and the $\xi^{(\gamma)}$ iteratively within a histogram
extrapolation framework, such as to simultaneously satisfy both a
generalized lever rule {\em and} equality of the probabilities of
occurrence of the phases, i.e.

\setcounter{abc}{1} 
\bea 
\label{eq:methoda} 
x^{(0)}_i &=&\sum_{\gamma=1}^m \xi^{(\gamma)}x^{(\gamma)}_i,\\ \addtocounter{abc}{1}
\addtocounter{equation}{-1} 
\effe &=&0\:.
\label{eq:methodb} 
\eea 
\setcounter{abc}{0} 
In the first of these constraints, Eq.~(\ref{eq:methoda}), the ensemble
averaged concentrations $x^{(\gamma)}$ are assigned by averaging only
over configurations belonging to the respective phase\footnote{Since in the SGCE
the entire system fluctuates between phases, one utilizes
the value of the order parameter associated with any given configuration in order to
ascertain the corresponding phase label $\gamma$.}. The deviation of the weighted sum $\bar x_i\equiv \sum_\gamma
\xi^{(\gamma)}x_i^{(\gamma)}$ from the target $x^{(0)}_i$ is
conveniently quantified by a `cost':
\begin{equation}  
\Delta\equiv \sum_{i=1}^3 \mid\bar x_i-x^{(0)}_i \mid  \;.  
\label{eq:costfn} 
\end{equation}  
In the second constraint, Eq.~(\ref{eq:methodb}),
\begin{equation}
\effe\equiv \sum_{\gamma=1}^m \left( \pi^{(\gamma)}-\frac{1}{m} \right)^2 
\end{equation}
provides a measure of the extent to which the probability of each phase
occuring, $\pi^{(\gamma)}$, is equal for each of the $m$ coexisting
phases. Imposing this equality ensures that phase properties are
measured under conditions of equal pressure and hence that finite-size
errors in these properties are exponentially small in the system
volume~\cite{Borgs1992,Buzzacchi2006}. In practice the relative
probabilities of the phases $\pi^{(\gamma)}$ is measured from the
individual peak weights of the order parameter pdf.

The determination of $\{\mu_i\}$ and $\xi^{(\gamma)}$ such as to
satisfy Eqs.~(\ref{eq:methoda}) and (\ref{eq:methodb}) proceeds
iteratively thus:

\begin{enumerate}

\item Guess initial values of the $\xi^{(\gamma)}$
corresponding to the chosen overall composition. Usually if starting
near the boundary of the coexistence region, $\xi^{(\gamma)}$ for the
incipient phase will be close to zero.

\item Tune $\{\mu_i\}$ (within the histogram extrapolation scheme) such as to minimize $\Delta$.

\item Measure the corresponding value of $\effe$.

\item if $\effe< {\rm tolerance}$, finish, otherwise vary
$\xi^{(\gamma)}$ (within the histogram extrapolation scheme) and repeat
from step 2.

\end{enumerate}

For the common case of two-phase coexistence, $\xi^{(2)}=1-\xi^{(1)}$ and
the minimization of $\effe$ with respect to variations in
$\xi^{(\gamma)}$ is a one-parameter minimization which is easily
automated using standard algorithms such as the ``Brent'' routine
described in Numerical Recipes~\cite{Numericalrecipes}.  In step $2$ the
minimization of $\Delta$ with respect to variations in $\{\mu_i\}$ is
most readily achieved~\cite{Wilding2002d} using the following simple
iterative scheme for $\{\mu_i\}$:

\begin{equation}
\mu^\prime_i=\mu_i+\alpha\ln\left( \frac{x_i^{(0)}} {\bar x_i}\right)\;.
\label{eq:update} 
\end{equation}
This update is applied simultaneously to each of the set of chemical
potentials $\{\mu_i\}$, and thereafter the set is shifted so that
$\mu_1=0$, where species $1$ is the chosen reference species. The
quantity $0<\alpha<1$ appearing in Eq.~(\ref{eq:update}) is a damping
factor, the value of which may be tuned to optimize the rate of
convergence. In order to ensure that finite-size effects are
exponentially small in the system size it is necessary to iterate to a
high tolerance in step $4$; we chose this to be $10^{-13}$.

The values of $\xi^{(\gamma)}$ resulting from the application of the
above procedure are the desired phase fractions at the nominated
$\{x^{(0)}\}$. Properties of the individual phases such as their average
number density, energy and concentrations are obtainable by
accumulating, for each phase separately, distributions of the respective
quantity. 

Having satisfied the above procedure for the initial coexistence state
point, one can track the path of interest deeper into the two phase
region with the aid of histogram extrapolation. One simply repeats the
procedure but with target concentrations that correspond to a state
point somewhat further along the path. Once the corresponding chemical
potentials have been estimated, a new SGCE simulation will provide new
data which in turn can be extrapolated yet further along the path. In
this way one steps along the path obtaining the coexistence properties
as one goes.

\subsubsection{Bootstrapping in the two phase region}
\label{sec:boot2phase}

If the composition space pathway of interests terminates on an axis of
the ternary phase diagram (ie. coincides with the limit of a binary
system) which furthermore, lies within a region of two phase
coexistence, the tracking procedure can be bootstrapped straightforwardly
at this state point. The strategy is first to eliminate the third
component by setting its chemical potential to a large negative value.
Thereafter, tuning the difference in chemical potential between the
remaining components allows one to scan along the coexistence tie line
which itself necessarily coincides with the axis of the phase diagram
(cf. Fig.~\ref{fig:schem_trian}). To achieve this one first links the
phases via Multicanonical preweighting (see
Refs.~\cite{Wilding1995,Bruce2003} for strategies that achieve this),
before applying the methodology outlined above to determine the chemical
potentials and $\xi^{(\gamma)}$ values corresponding to the terminal
point on the pathway. Thereafter one increases the chemical potential of
the third component so that it has a small but finite concentration,
which then facilitates tracking the path of interest via histogram
extrapolation, as outlined above.

\section{Illustrative results}

We illustrate our procedure with results for a ternary Lennard Jones
mixture, defined by

\begin{equation}
u_{ij}=4\epsilon_{ij}\left[ \left(\frac{ \sigma_{ij} }{ r_{ij} }\right)^{12}-\left(\frac{\sigma_{ij}}{r_{ij}}\right)^{6} \right]\;,
\end{equation}
where
\begin{equation}
\sigma_{ij}=\frac{\sigma_{ii}+\sigma_{jj}}{2};\hspace*{1cm}\epsilon_{ij}=\sqrt{\epsilon_{ii}\epsilon_{jj}}\;;
\end{equation}
with
\begin{eqnarray*}
\sigma_{22}/\sigma_{11} &= 0.75;\hspace*{6mm} & \sigma_{33}/\sigma_{11} = 0.5\;; \\
\epsilon_{22}/\epsilon_{11} &= 0.75;\hspace*{6mm} &\epsilon_{33}/\epsilon_{11}= 0.5\;.\\
\end{eqnarray*}
The potential was truncated at $r_{ij}=3.0\sigma_{ij}$ and no tail correction was applied.

We have studied the phase behaviour of this model for a system of
$N=256$ particles at the reduced temperature
$T^\star=k_BT/\epsilon_{11}=1.0$ and the reduced pressure
$p^\star=p\sigma_{11}/\epsilon_{11}=0.2$. A similar system has been
studied previously by Tsang et al \cite{Tsang1995},
Escobedo\cite{Escobedo1999} and Errington\cite{Errington2005}, although
unlike the present work these studies utilized a correction for the
potential truncation, so their results are not directly comparable with ours.

The path we chose to study is one in which we fixed $x_1^{(0)}=0.41$ and
varied $x_2$ in the range $[0,0.59]$. The choice for $x_2^{(0)}$ then
parameterizes the location along the path. The tracking procedure was
bootstrapped at $x_2^{(0)}=0$ which corresponds to a binary system comprising
components $1$ and $3$. As also observed by Tsang {\em et al}
\cite{Tsang1995}, vapor-liquid coexistence occurs in this limit. We
therefore employed the bootstrapping approach outlined in
sec.~\ref{sec:boot2phase} which yielded the value $\xi^{(2)}=0.492(2)$ for
the liquid phase fraction. Once bootstrapped, we performed
a further run in which a small but finite value of $\exp(\mu_2/T)$ was used in
order to generate fluctuations in $x_2^{(0)}$, thus facilitating tracking of
the path of interest in the manner described in Sec.~\ref{sec:tracking}.

\begin{figure}[h]
\includegraphics[type=pdf,ext=.pdf,read=.pdf,width=0.90\columnwidth,clip=true]{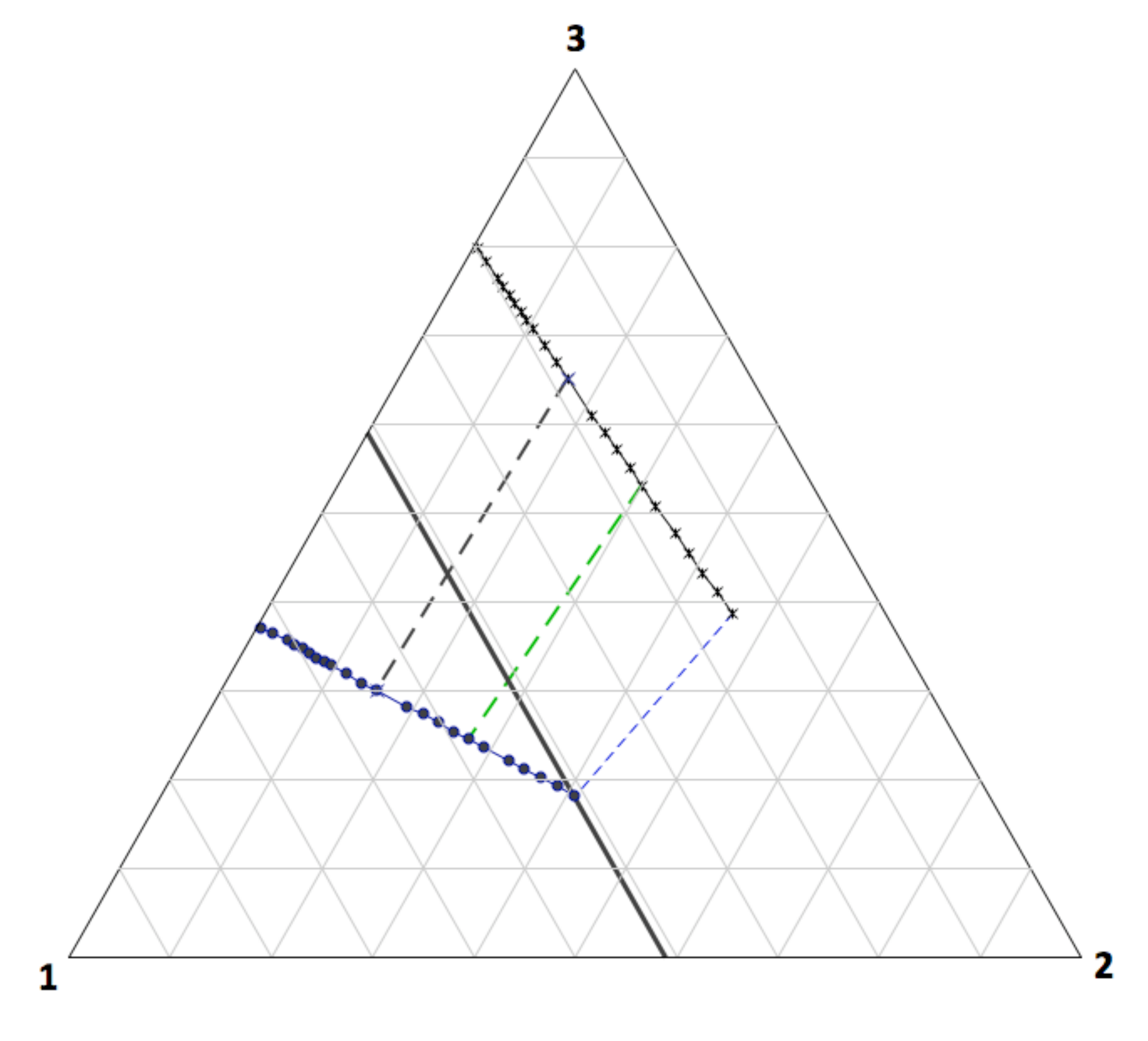}\\
\caption{The thick black line indicates the path in overall composition
tracked in the simulations. Symbols denote the compositions of the
coexisting phases (crosses are for the gas; filled circles are liquid),
observed along this path. Statistical uncertainties do not exceed the
symbol sizes. A selection of representative tie lines are
also shown (dashed lines). At $x_2^{0)}=0.409(1)$, the path leaves the region of phase
coexistence.}
\label{fig:LJPD} 
\end{figure}

\begin{figure}[h]
\includegraphics[type=pdf,ext=.pdf,read=.pdf,width=0.95\columnwidth,clip=true]{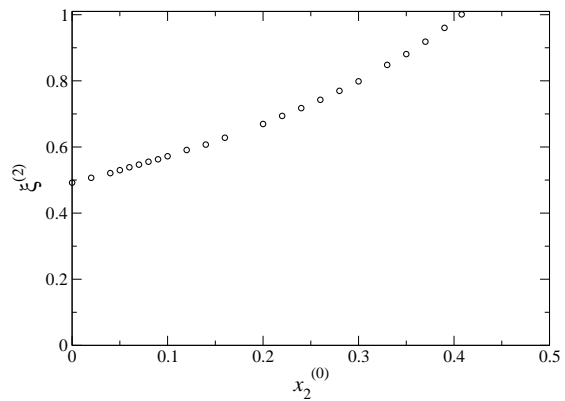}\\
\caption{Measured value of the liquid phase fraction $\xi^{(2)}$ as a function of the path parameter $x_2^{(0)}$}
\label{fig:xi} 
\end{figure}

Our results for the phase behaviour are summarized in
fig.~\ref{fig:LJPD} which shows the path in overall composition (thick
black line) as well as the compositions of the coexisting phases
observed along this path (symbols). The corresponding measurements of
$\xi^{(2)}$, the fraction of liquid, as a function of the path parameter
$x_2^{(0)}$ are shown in Fig.~\ref{fig:xi}.

\section{Conclusions}

In summary we have described a computational procedure for accurately
and directly determining the phase behaviour that occurs along a path in
the space of the {\em overall} composition of a ternary fluid mixture.
For each state point along the path the method yields estimates for the
concentrations of the coexisting phases whose tie line intersects that
point, together with the phase fractions. Since finite-size corrections
are exponentially small in the system size, the results are highly
accurate even when the path of interest touches the coexistence boundary
where the fraction of one phase vanishes. The simulation scheme, which
is based on an extended sampling scheme in the semigrand canonical
ensemble, uses histogram reweighting to track the path of interest. The
only requirement is to bootstrap the method by determining the set of
chemical potential differences that yield the overall concentrations
corresponding to some point on the path. We have outlined how this can
be achieved for pathway points corresponding to a single phase region of
the phase diagram, or within a coexistence region in the limit in which
one component vanishes, ie. for a binary system.

Our method is suitable for both fluid-fluid and fluid-solid phase
coexistence, although in the latter case it needs to be combined with a
sampling scheme such as phase switch Monte Carlo which allows  the
configuration spaces of the two phase to be connected by a trouble free
sampling path~\cite{Wilding2000,Wilding2009}. By virtue of the
flexibility of histogram extrapolation, it is also readily
generalizable to paths that involve simultaneous variation in
composition and an external field (temperature or pressure). In future
work we plan to demonstrate its utility for tracking such paths in the
contexts of mixtures of four or more components for which compact
phase diagrams do not exist.

\bibliography{/Users/pysnbw/Documents/Papers}

\end{document}